\documentclass[acus]{JAC2003}
\usepackage{graphicx}
\begin{document}

\thispagestyle{empty}

\onecolumn

\begin{flushright}
{\large
SLAC--PUB--9810\\
May 2003\\}
\end{flushright}

\vspace{.8cm}

\begin{center}

{\LARGE\bf
Lattice with Smaller Momentum Compaction Factor\\
for PEP-II High Energy Ring~\footnote
{\normalsize
{Work supported by Department of Energy contract  DE--AC03--76SF00515.}}}

\vspace{1cm}

\large{
Y.~Cai, M.H.R.~Donald, Y.~Nosochkov\\
Stanford Linear Accelerator Center, Stanford University,
Stanford, CA 94309}

\end{center}

\vfill

\begin{center}
{\LARGE\bf
Abstract }
\end{center}

\begin{quote}
\large{
At present, the PEP-II bunch length and vertical beta function $\beta_y^*$
at the Interaction Point (IP) are about of the same size.  To increase
luminosity, it is planned to gradually reduce $\beta_y^*$.  For the maximum
effect, bunch length has to be also reduced along with $\beta_y^*$ to
minimize luminosity loss caused by the hourglass effect at IP.  One of the
methods to achieve a smaller bunch length is to reduce momentum compaction
factor.  This paper discusses a lattice option for the High Energy Ring,
where the nominal $60^\circ$ cells in four arcs are replaced by $90^\circ$
cells to reduce momentum compaction factor by 30\% and bunch length by
16\%.  The increased focusing in $90^\circ$ cells results in 40\% stronger
arc quadrupoles and 150\% stronger arc sextupoles due to reduced dispersion
and larger chromaticity.  Tracking simulations predict that dynamic
aperture for this lattice will be $\geq10$ times the {\it rms} size of a
fully coupled beam for a horizontal emittance of 30~nm and
$\beta_y^*\!=\!1$~cm.  The lattice modification and results of simulations
are presented.
}
\end{quote}

\vfill

\begin{center}
\large{
{\it Presented at the 2003 Particle Accelerator Conference
(PAC 03)\\
Portland, Oregon, May 12--16, 2003}
} \\
\end{center}

\newpage

\pagenumbering{arabic}
\pagestyle{plain}

\twocolumn

\title
{LATTICE WITH SMALLER MOMENTUM COMPACTION FACTOR\\
FOR PEP-II HIGH ENERGY RING~\thanks
{Work supported by Department of Energy contract 
DE--AC03--76SF00515.}\vspace{-4mm}}

\author{Y.~Cai, M.H.R.~Donald, Y.~Nosochkov,
SLAC, Menlo Park, CA 94025, USA}

\maketitle

\begin{abstract} 

At present, the PEP-II bunch length and vertical beta function $\beta_y^*$
at the Interaction Point (IP) are about of the same size.  To increase
luminosity, it is planned to gradually reduce $\beta_y^*$.  For the maximum
effect, bunch length has to be also reduced along with $\beta_y^*$ to
minimize luminosity loss caused by the hourglass effect at IP.  One of the
methods to achieve a smaller bunch length is to reduce momentum compaction
factor.  This paper discusses a lattice option for the High Energy Ring,
where the nominal $60^\circ$ cells in four arcs are replaced by $90^\circ$
cells to reduce momentum compaction factor by 30\% and bunch length by
16\%.  The increased focusing in $90^\circ$ cells results in 40\% stronger
arc quadrupoles and 150\% stronger arc sextupoles due to reduced dispersion
and larger chromaticity.  Tracking simulations predict that dynamic
aperture for this lattice will be $\geq10$ times the {\it rms} size of a
fully coupled beam for a horizontal emittance of 30~nm and
$\beta_y^*\!=\!1$~cm.  The lattice modification and results of simulations
are presented.

\end{abstract}

\section{INTRODUCTION}

One of the methods to increase luminosity at PEP-II~\cite{cdr} is to reduce
a vertical beta function $\beta_y^*$ at the Interaction Point (IP).  The
current plan is to reduce $\beta_y^*$ from the present value of 12.5~mm to
9~mm this year, and to $\sim\,$5~mm within the next few years.

\begin{figure}[tb]
\centering
\includegraphics*[width=75mm]{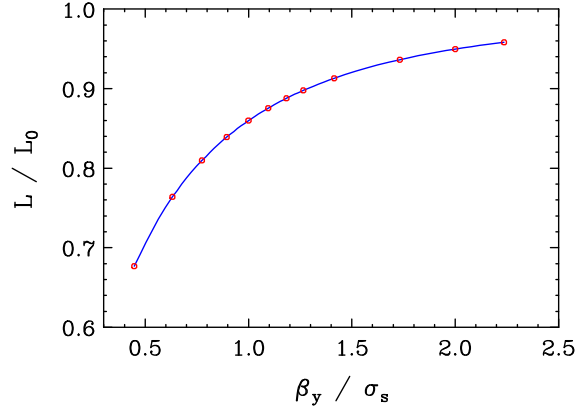}
\vspace{-3mm}
\caption{Luminosity reduction factor due to the hourglass effect
for flat beams.}
\label{fig:hglass}
\vspace{-2mm}
\end{figure}

Due to a finite bunch length $\sigma_s$, particle interactions occur over
distance $-\sigma_s/2\!<\!s\!<\!\sigma_s/2$ from IP.  Because of angular
divergence $\propto\!1/\sqrt{\beta_y^*}$, beam size increases with distance
$s$ from IP according to:
$\sigma_y(s)\!=\!\sigma_y^*\sqrt{1+s^2/{\beta_y^*}^2}$.  As a result,
contribution to luminosity is gradually reduced with distance from the beam
waist at IP.  This so-called ``hourglass'' effect can be analytically
estimated and translated into a luminosity reduction factor due to a finite
bunch length~\cite{furman}.  For flat beams with equal beam size and
emittance, this factor depends only on one parameter $\beta_y^*/\sigma_s$
and is shown in Fig.~\ref{fig:hglass}.

At present, the bunch length and $\beta_y^*$ at PEP-II are about of the
same size.  According to Fig.~\ref{fig:hglass}, this corresponds to 14\% of
luminosity loss due to the hourglass effect.  If $\beta_y^*$ is reduced
from the current 12.5~mm to 9~mm and then to 5~mm without changing
$\sigma_s$, luminosity loss would increase to 21\% and 35\%, respectively.
One can conclude, therefore, that for maximum PEP-II luminosity at lower
$\beta_y^*$, bunch length has to be reduced as well.

Among other parameters, the equilibrium bunch length depends on the total
accelerating rf-voltage $V$, momentum compaction factor $\alpha$ and 
bending radius $\rho$ as
\begin{equation}
\sigma_s \propto \sqrt{\frac{\alpha}{V} \cdot
\frac{\langle\left|1/\rho^3\right|\rangle}{\langle1/\rho^2\rangle}},
\label{eqn:sigmas}
\end{equation}
where $\langle\rangle$ denote an average in the machine~\cite{wiedmn}.
The PEP-II upgrade to increase rf-voltage for a smaller bunch length
is being implemented.  However, the bunch length is a relatively
slow function of $V$, therefore many rf-cavities would be needed for
a large reduction of $\beta_y^*$.  To help reduce the bunch length, a
reduction of momentum compaction factor may be considered.

The momentum compaction factor is defined by dispersion function $\eta_x$
and bending radius $\rho$ according to
\begin{equation}
\alpha = \langle \frac{\eta_x}{\rho} \rangle,
\label{eqn:alpha}
\end{equation}
where $\eta_x$ depends on $\rho$ and quadrupole focusing.  A change of
bending properties or magnet locations is not considered in this paper
since it would require a modification of machine geometry.  Therefore, for
a fixed bending, a smaller momentum compaction factor could be achieved by
reducing the average dispersion in bends by means of a stronger quadrupole
focusing.  Such optics modification is discussed below for the PEP-II High
Energy Ring (HER) with $\beta_x^*/\beta_y^*\!=\!50/1$~cm.

\section{LATTICE MODIFICATION}

\begin{figure}[tb]
\centering
\includegraphics*[width=65mm]{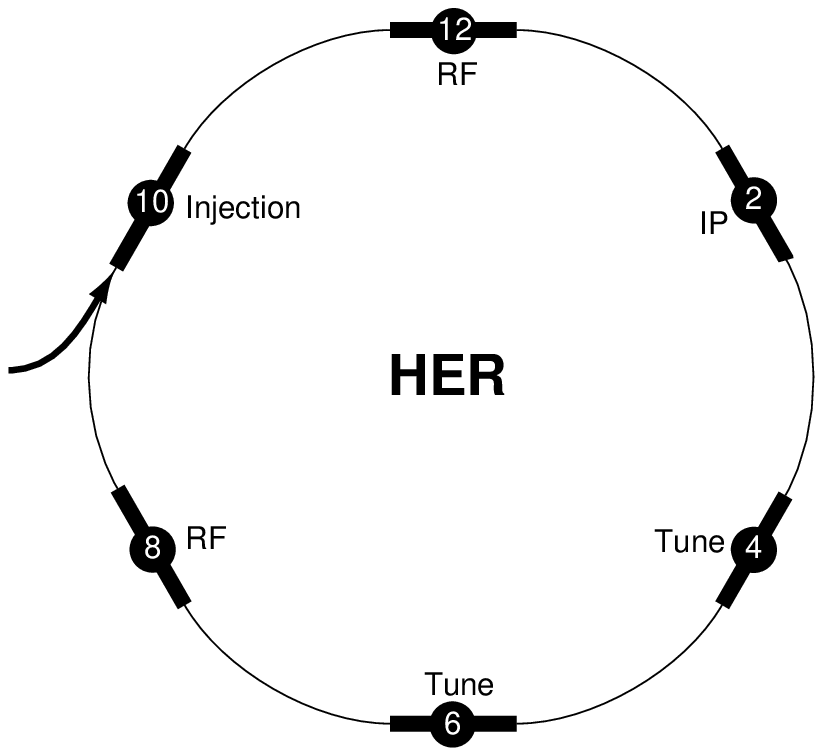}
\vspace{-0mm}
\caption{Top view of the High Energy Ring.}
\label{fig:ring}
\vspace{-1mm}
\end{figure}

\begin{figure}[tb]
\centering
\includegraphics*[width=82mm]{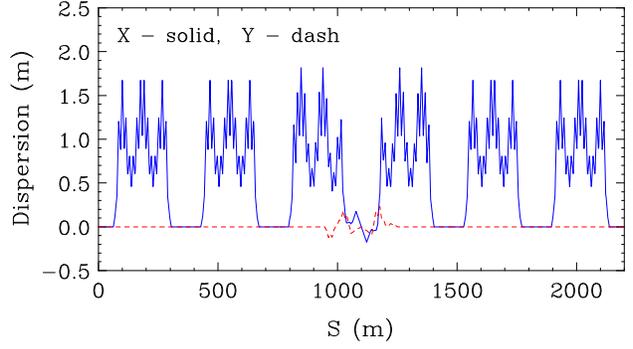}
\vspace{-7mm}
\caption{Dispersion in the nominal HER.}
\label{fig:disp60}
\vspace{-2mm}
\end{figure}

Layout of the HER is shown in Fig.~\ref{fig:ring}.  The lattice consists
of six arcs with periodic $60^\circ$ cells and six straight sections with
various matched optics for the Interaction Region (IR), injection,
rf-cavities, and tune and coupling correction.  The HER nominal dispersion
function is shown in Fig.~\ref{fig:disp60}, where IP is in the middle at
$s\!\approx\!1100$~m.  Modulation of $\eta_x$ in the four arcs farthest
from IR is introduced to increase the HER horizontal emittance to
48~nm, while in the two arcs near IR it is caused by special $\beta$ bumps
for the IR sextupoles.  Because in the four arcs this perturbation is a
free betatron motion around the periodic $\eta_x$, it does not change the
average dispersion $\langle\eta_x\rangle$ and bunch length, but increases
$\langle\eta_x^2\rangle$ for a higher emittance. In the straight sections,
dispersion is canceled by dispersion suppressors.

The most contribution to momentum compaction factor in HER comes
from dispersion in the arcs.  A simple way to reduce $\langle\eta_x\rangle$
is to increase phase advance in the periodic arc cells.  The effect of
phase advance per cell $\mu_c$ can be estimated using a thin lens
approximation.  This method gives the following well-known equations for
the extreme ($\pm$) values of $\beta$ and $\eta_x$, and the quadrupole
integrated strength $K_{1}L$ in the arc FODO cell:
\begin{eqnarray}
\beta^\pm = L_c \, \frac{1\pm\sin(\mu_c/2)}{\sin\mu_c},
\;\;\;
\label{eqn:betpm} \\
\eta_x^\pm = \frac{L_c^2}{8\rho} \cdot \frac{2\pm\sin(\mu_c/2)}
{\sin^2(\mu_c/2)},
\label{eqn:dxpm} \\
K_{1}L = \frac{4\sin(\mu_c/2)}{L_c},
\;\;\;\;\;\;\;\;\;\,
\label{eqn:kl}
\end{eqnarray}
where $L_c$ is a cell length.  For an estimate of the average values of
$\beta$ and $\eta_x$ in the arcs, one could use the following approximation:
\begin{eqnarray}
\langle\beta\rangle \,\approx\, \frac{\beta^+ + \beta^-}{2} = 
\frac{L_c}{\sin\mu_c},
\;\;\;\;\;\;\;\;\;\;\;\,
\label{eqn:betav} \\
\langle\eta_x\rangle \,\approx\, \frac{\eta_x^+ + \eta_x^-}{2} =
\frac{L_c^2}{4\rho\sin^2(\mu_c/2)}.
\label{eqn:dxav}
\end{eqnarray}

Below, a modification of HER optics for a lower momentum compaction factor is
considered, where phase advance in the four arcs farthest from IR is increased
from $60^\circ$ to $90^\circ$ per cell.  The other two arcs contain some of the
IR sextupoles and skew quadrupoles to compensate the detector solenoid and
non-linear chromaticity.  In order to maintain the original IR optics and local
correction, these two arcs were not changed.

Since the maximum $\beta$ functions are about the same in $60^\circ$ and
$90^\circ$ cells, physical aperture acceptance will not be reduced by this
modification.  As in $60^\circ$ optics, the $90^\circ$ cells naturally
provide $-I$ transformation between arc sextupoles to help compensate the
third order sextupole aberrations.  Also, the first order chromatic
perturbation of $\beta$ function is naturally suppressed in $90^\circ$
lattice.

According to Eqn.~\ref{eqn:dxav}, the average dispersion in $90^\circ$
arc is reduced by a factor of 2 compared to $60^\circ$ lattice.
Consequently, the momentum compaction factor in four $90^\circ$ and two
$60^\circ$ arcs is reduced by a factor of $\frac{2}{3}$ compared to the
$60^\circ$ value.  From Eqn.~\ref{eqn:betpm}--\ref{eqn:kl}, disadvantages
of $90^\circ$ cells are a factor of $\sqrt{2}$ stronger quadrupoles, a
factor of $\sqrt{3}$ larger linear chromaticity per cell and a factor of
$2\sqrt{2}$ stronger sextupoles ($K_{2}\!\propto\!K_{1}/\eta_x$).

To maintain the original non-dispersive optics in the injection, tuning and
rf-cavity sections, quadrupole focusing in dispersion suppressors designed
for $60^\circ$ arcs was appropriately adjusted to match the straight
sections to the new $\beta$ functions and reduced dispersion in
$90^\circ$ arcs.  One complication was related to the original design of
arc sextupoles, where each of the four arcs has 12 SF and 12 SD sextupoles
to correct linear chromaticity.  Ideally, the same family sextupoles should
have identical lattice functions to minimize residual sextupole
aberrations.  But in the HER, 2 SF and 2 SD sextupoles in each arc are
extended into the dispersion suppressors which have different optics
compared to the arcs.  In the original $60^\circ$ design, lattice functions
at the above 4 sextupoles were made reasonably close to the periodic values
in the arcs.  It has been found particularly important to keep this
property in the $90^\circ$ modification as well.  It was verified that a
large change of $\beta$ functions at these sextupoles could reduce dynamic
aperture to unacceptable level.  This is caused by an increase of the
third order sextupole geometric aberrations if they are not sufficiently
compensated due to breakdown of optical periodicity and $-I$ transformation
at the 4 sextupoles.

\begin{figure}[tb]
\centering
\includegraphics*[width=82mm]{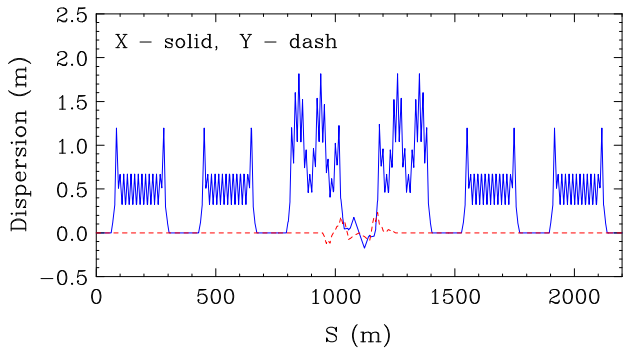}
\vspace{-7mm}
\caption{Dispersion in HER with reduced $\alpha$.}
\label{fig:disp90}
\vspace{-2mm}
\end{figure}

The resultant dispersion in HER with four $90^\circ$ arcs is shown in
Fig.~\ref{fig:disp90}.  In this option, a periodic dispersion without
modulation is used in the $90^\circ$ arcs, while dispersion in the two arcs
near IR is not changed.  Some of the HER global parameters for the original
$60^\circ$ and modified $90^\circ$ lattice with
$\beta_x^*/\beta_y^*\!=\!50/1$~cm are shown in Table~1, where the total
voltage of $V\!=\!14$~MV was used.

\begin{table}[tb]
\small
\begin{center}
\vspace{-1mm}
\caption{HER parameters for $60^\circ$ and $90^\circ$ lattice.}
\medskip
\begin{tabular}{|l|c|c|c|c|c|}
\hline
\boldmath{$\mu_c$} & \boldmath{$\alpha$} & \boldmath{$\epsilon_x$} &  
\boldmath{$\nu_x\,/\,\nu_y$} & \boldmath{$\nu_s$} & 
\boldmath{$\xi_x\,/\,\xi_y$} \\
 & \boldmath{$[10^{-3}]$} & \textbf{[nm]} & & & \\
\hline
$60^\circ$ & 2.41 & 48 & 24.569\,/\,23.639 & 0.045 & -44\,/\,-71 \\
$90^\circ$ & 1.69 & 30 & 28.569\,/\,29.639 & 0.038 & -56\,/\,-81 \\
\hline
\end{tabular}
\label{tab:param}
\end{center}
\vspace{-5mm}
\end{table}

Momentum compaction factor is reduced by 30\% in the $90^\circ$ modification,
therefore the bunch length is expected to decrease by 16\%.  The reduced
dispersion in $90^\circ$ arcs results in a smaller horizontal emittance
$\epsilon_x$ in this option.  A modulation of $\eta_x$ may be introduced to
increase the emittance.  For the same rf-voltage, synchrotron tune $\nu_s$
is also reduced by 16\% since it scales as $\sqrt{\alpha\,V}$.  If the
voltage is increased for a smaller bunch length, $\nu_s$ could be restored.

Naturally, the stronger quadrupole focusing in $90^\circ$ arcs increases
the HER betatron tune $\nu_x/\nu_y$ and linear chromaticity
$\xi_x/\xi_y$.  Quadrupole strength increases by 42\% in the $90^\circ$
arcs, and the SF, SD sextupoles become stronger by a factor of 2.3 and 2.5,
respectively, compared to $60^\circ$ design.  The large increase in
strength may require an upgrade for some of these magnets.

\begin{figure}[tb]
\centering
\includegraphics*[width=82mm]{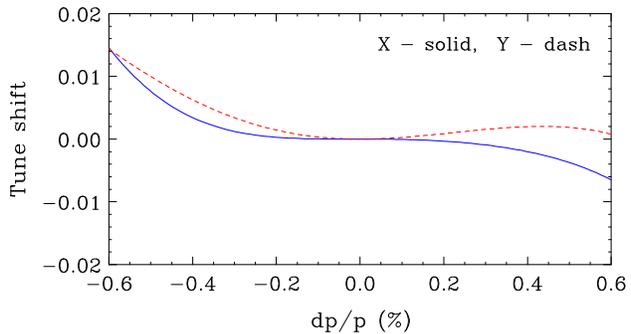}
\vspace{-6mm}
\caption{Tune shift vs. $\frac{\Delta p}{p}$ in HER with reduced $\alpha$.}
\label{fig:tune90}
\vspace{-3mm}
\end{figure}

Optics and compensation schemes of the Interaction Region have not been
changed in this modification.  The IR sextupoles provide correction of the
non-linear chromaticity generated in the final quadrupole doublets near IP.
It has been important to verify that compensation of non-linear
chromaticity has not been affected by arc modifications.  Indeed,
calculation of betatron tune and $\beta^*$ in the $90^\circ$ lattice versus
relative momentum deviation $\frac{\Delta p}{p}$ showed a negligible change
of non-linear chromaticity compared to the original optics.  This confirms
that the IR chromaticity correction is, indeed, local.  Tune shift in the
modified HER is shown in Fig.~\ref{fig:tune90} for the range of
$-10\sigma_p\!<\!\frac{\Delta p}{p}\!<\!10\sigma_p$, where $\sigma_p$ is
the {\it rms} relative energy spread in the beam, and linear chromaticity
is set to zero.

\begin{figure}[htb]
\centering
\includegraphics*[width=60mm]{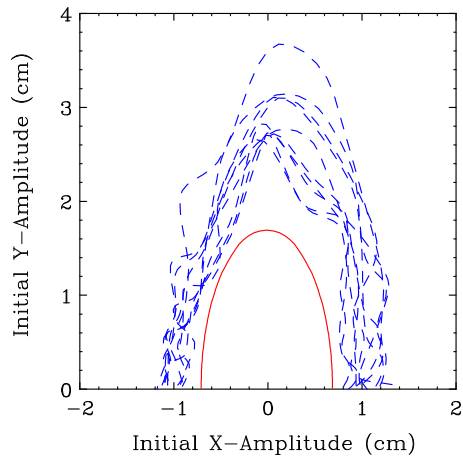}
\vspace{-3mm}
\caption{Dynamic aperture in HER with reduced $\alpha$.}
\label{fig:aper90}
\vspace{-2mm}
\end{figure}

Finally, tracking simulations have been performed to verify dynamic
aperture for the HER with $90^\circ$ arcs and $\beta_y^*\!=\!1$~cm.
Simulations have been done using LEGO code~\cite{lego} for 10 different
combinations of random field and alignment errors, and $\pm8\sigma_p$
synchrotron oscillations.  Compensation of beam orbit, linear chromaticity,
coupling and tune were simulated in LEGO prior to tracking.  The resultant
dynamic aperture at the injection point is shown in Fig.~\ref{fig:aper90},
where the 10 dash lines represent different error settings.  The area
inside a dash line corresponds to a particle stable motion.  This dynamic
aperture exceeds $10\sigma$ depicted by a solid line, where $\sigma$ is the
{\it rms} size of a fully coupled beam at injection with
$\epsilon_x\!=\!30$~nm and $\epsilon_y\!=\!\epsilon_x/2$.  This dynamic
aperture should be sufficient for beam operation.  We conclude, therefore,
that $90^\circ$ optics in HER for a lower momentum compaction factor
may be considered as an option for a shorter bunch length.

\section{CONCLUSION}

It has been shown that momentum compaction factor in HER can be reduced by
30\% by increasing phase advance per cell from $60^\circ$ to $90^\circ$ in
four arcs.  The resultant dynamic aperture exceeds $10\sigma$ and is
considered adequate.  The expected reduction of bunch length in this option
is 16\%.


\end{document}